\documentclass[conference]{IEEEtran}
\IEEEoverridecommandlockouts
\usepackage{cite}
\usepackage{amsmath,amssymb,amsfonts}
\usepackage{algorithmic}
\usepackage{graphicx}
\usepackage{subcaption}
\usepackage{textcomp}
\usepackage{xcolor}
\usepackage{comment}
\usepackage{booktabs}
\usepackage{comment}
\usepackage{url}
\usepackage[english]{babel}
\usepackage[noabbrev]{cleveref}

\def\BibTeX{{\rm B\kern-.05em{\sc i\kern-.025em b}\kern-.08em
    T\kern-.1667em\lower.7ex\hbox{E}\kern-.125emX}}
\begin{document}

\title{Privacy-Preserving Data Sharing in Agriculture: Enforcing Policy Rules for Secure and Confidential Data Synthesis}

\makeatletter
\newcommand{\linebreakand}{%
  \end{@IEEEauthorhalign}
  \hfill\mbox{}\par
  \mbox{}\hfill\begin{@IEEEauthorhalign}
}
\makeatother

\author{\IEEEauthorblockN{Anantaa Kotal}
\IEEEauthorblockA{\textit{Dept. of C.S.E.E.} \\
\textit{University of Maryland, Baltimore County}\\
Baltimore, USA \\
anantak1@umbc.edu}
\and
\IEEEauthorblockN{Lavanya Elluri}
\IEEEauthorblockA{\textit{Dept. of Computer Information Systems} \\
\textit{Texas A\&M University - Central Texas}\\
Killeen, USA  \\
elluri@tamuct.edu}
\and
\IEEEauthorblockN{Deepti Gupta}
\IEEEauthorblockA{\textit{ Dept. of Computer Information Systems } \\
\textit{Texas A\&M University - Central Texas}\\
Killeen, USA \\
d.gupta@tamuct.edu }
\linebreakand 
\IEEEauthorblockN{Varun Mandalapu }
\IEEEauthorblockA{\textit{ Dept. of Information Systems } \\
\textit{ University of Maryland, Baltimore County }\\
Baltimore, USA \\
varunm1@umbc.edu}
\and
\IEEEauthorblockN{Anupam Joshi}
\IEEEauthorblockA{\textit{Dept. of C.S.E.E.} \\
\textit{University of Maryland, Baltimore County}\\
Baltimore, USA \\
joshi@umbc.edu}
}

\IEEEoverridecommandlockouts
\IEEEpubid{\makebox[\columnwidth]{979-8-3503-2445-7/23/\$31.00~\copyright2023 IEEE \hfill}
\hspace{\columnsep}\makebox[\columnwidth]{ }}
\maketitle
\begin{abstract}
Big Data empowers the farming community with the information needed to optimize resource usage, increase productivity, and enhance the sustainability of agricultural practices. The use of Big Data in farming requires the collection and analysis of data from various sources such as sensors, satellites, and farmer surveys. While Big Data can provide the farming community with valuable insights and improve efficiency, there is significant concern regarding the security of this data as well as the privacy of the participants. Privacy regulations, such as the European Union's General Data Protection Regulation (GDPR), the EU Code of Conduct on agricultural data sharing by contractual agreement, and the proposed EU AI law, have been created to address the issue of data privacy and provide specific guidelines on when and how data can be shared between organizations. To make confidential agricultural data widely available for Big Data analysis without violating the privacy of the data subjects, we consider privacy-preserving methods of data sharing in agriculture. Synthetic data that retains the statistical properties of the original data but does not include actual individuals' information provides a suitable alternative to sharing sensitive datasets. Deep learning-based synthetic data generation has been proposed for privacy-preserving data sharing. However, there is a lack of compliance with documented data privacy policies in such privacy-preserving efforts. In this study, we propose a novel framework for enforcing privacy policy rules in privacy-preserving data generation algorithms. We explore several available agricultural codes of conduct, extract knowledge related to the privacy constraints in data, and use the extracted knowledge to define privacy bounds in a privacy-preserving generative model. We use our framework to generate synthetic agricultural data and present experimental results that demonstrate the utility of the synthetic dataset in downstream tasks. We also show that our framework can evade potential threats, such as re-identification and linkage issues, and secure data based on applicable regulatory policy rules.

\end{abstract}


\section{Introduction}
According to the Food and Agriculture Organization (FAO) \cite{FAO}, food production needs to be raised by 70\% by 2050 to feed the projected population of 9.6 billion by 2050. To meet the growing needs of an expanding population, the farming community needs efficient ways to enhance agricultural productivity, optimize resource utilization, and implement sustainable farming practices. Technology plays an essential role in meeting these goals. Specifically, the use of Big Data in agriculture enables us to model large volumes of crowd-sourced data, sensor measurement data, and environmental information to understand emerging patterns in agriculture. This helps us stay ahead of modern challenges such as changing climate, farmland depletion, etc. 

However, sharing data from multiple sources raises privacy concerns. Privacy regulations, such as the European Union's General Data Protection Regulation (GDPR) \cite{EUGDPR}, have specific requirements on when and how such data can be shared. Even in the absence of specific regulations, individuals may have significant concerns about sharing their private information. We need privacy-preserving methods of sharing agricultural data that abide by policy regulations and encourage the farming community to participate in collective study.

The European Union's General Data Protection Regulation (GDPR) is one of the most comprehensive data protection regulations globally and has implications for agricultural data in EU member states. However, it does not specifically address data sharing in agriculture. The EU Code of Conduct on agricultural data sharing by contractual agreement \cite{EUCode} was created as an addendum to the GDPR. This code of conduct (Code) was designed to promote data-sharing leads in agriculture by setting transparent principles, clarifying responsibilities, and creating trust among partners. The EU A.I. Act proposed in April 2021 by the European Commission proposed as the first EU regulatory framework for AI. It is the first regulation of its kind to define and classify AI used in different applications according to the risk they pose to users. AI models that pose higher risks require stricter regulation. In this study, the EU Code of Conduct on agricultural data sharing by contractual agreement \cite{EUCode}, and the E.U. A.I. Act \cite{AIACT} have been identified as relevant privacy policies that can help us understand privacy concerns related to Big Data in agriculture.

Privacy in shared data has been discussed in prior work focusing on anonymization and encryption of data. However, these methods can be costly and not easily scalable. Privacy-preserving data generation refers to the process of creating new {\em synthetic} data that maintains privacy while retaining useful characteristics and statistical properties of the original data.  Privacy-preserving data generation is a robust way of protecting sensitive data while still making them useful for wide-scale sharing. Generative Adversarial Networks (GANs) are one of the well-known models for generating synthetic samples that can have the same distributional characteristics as the original data. Synthetic data generated using privacy-preserving versions of GAN have been shown to replace real data for statistical and analytical purposes while protecting sensitive information \cite{torkzadehmahani2019dp, xu2018dp, jordon2018pate, kotal2022privetab}. While previous studies have successfully generated privacy preserving of synthetic data, they do not consider the requirements of privacy regulations. Thus, there is a gap between data privacy as defined by the privacy policies and the privacy constraints in the privacy-preserving methods.

To create a secure and privacy-preserving version of agricultural data that aligns with the regulations outlined in policy frameworks, we present an innovative framework. This framework is designed to generate data while strictly adhering to the guidelines established in agricultural privacy regulations. By incorporating the definition of privacy from these regulations into our model, we establish privacy constraints that specifically address the nuances of shared agricultural data.

The resulting synthetic dataset serves as a privacy-preserving alternative to confidential data, ensuring compliance with the principles outlined in agricultural codes of conduct. Our proposed framework effectively mitigates a range of threats, including privacy leakage, re-identification, side-channel, linkage, and attribute inference attacks. Through extensive experimentation, our model demonstrates resilience against various privacy threat models. Importantly, it achieves this while maintaining both statistical similarity to the original data and practical utility in downstream tasks.

The remainder of this paper is organized as follows. Section~\ref{background} establishes the background and motivation for our work in the context of potential threats against privacy-preserving framework, privacy policy regulations, and privacy-preserving data generation efforts. Section~\ref{sec_framework} presents our framework for policy enforcement in privacy-preserving data generation through rule extraction from privacy regulation and attribute regulation in generative models. Section~\ref{results} presents the results of our experimental work on policy-enforced data sharing Agricultural data. Section~\ref{related} presents the literature review on privacy regulations and methods to secure agriculture data from potential threats. The conclusion and future work are shown in Section~\ref{conclusion}.

\section{Background and Motivation}
\label{background}

\subsection{Privacy Threat Models in Agriculture}

\label{threat}
\begin{figure}
\centering
\includegraphics[width=0.5\textwidth]{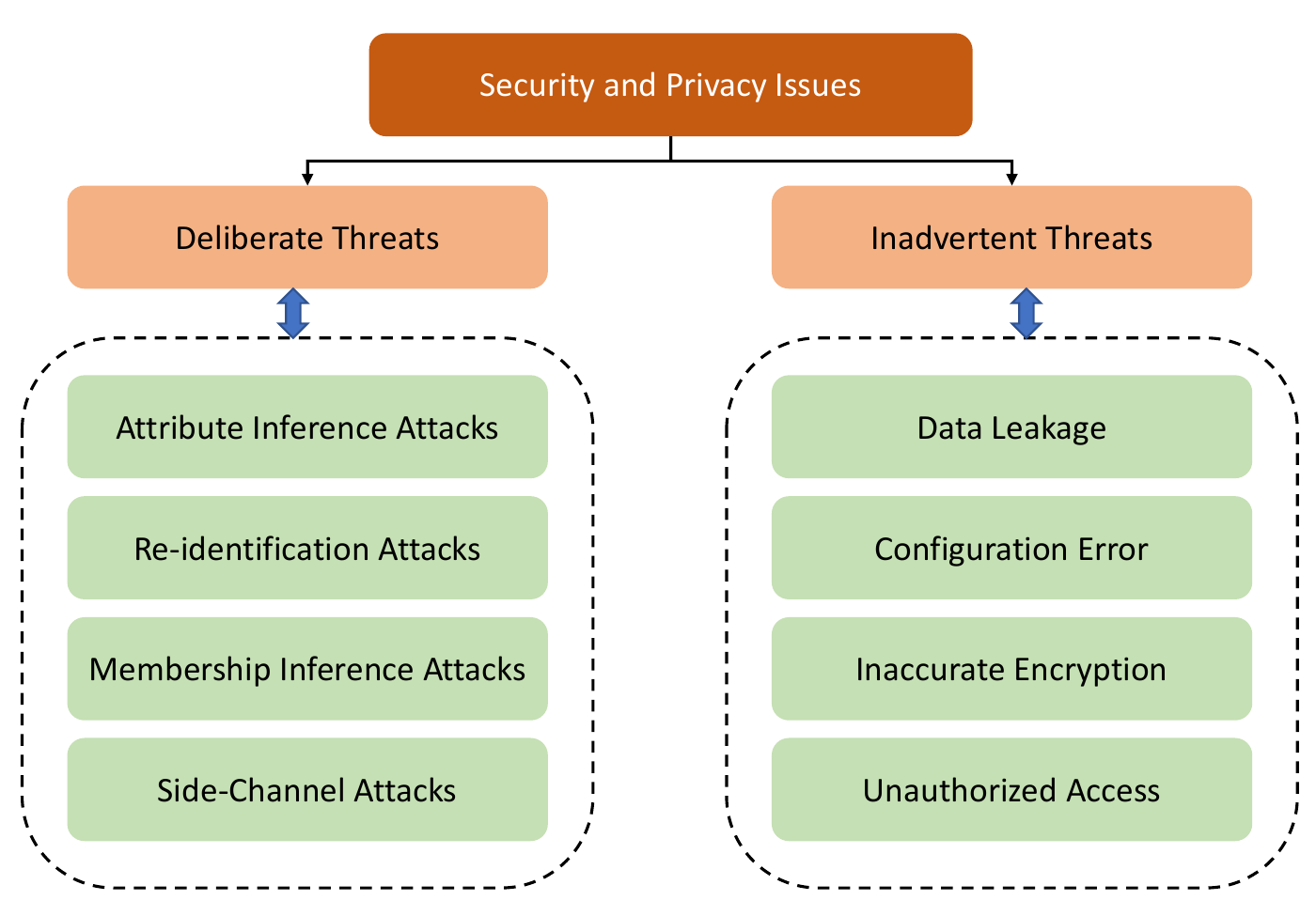}
\centering
\caption{Overview of Potential Threats.}
\label{fig:threats}
\end{figure}

In the agricultural domain, privacy-preserving mechanisms are designed to protect individuals' sensitive information while allowing data to be analyzed and utilized for various purposes. However, these mechanisms can also be vulnerable to specific attacks that attempt to compromise the privacy they aim to preserve. Here are some potential threats against privacy-preserving measures, where threats are categorized as intentional and unintentional threats. 

\subsubsection{Deliberate Threats}
\begin{itemize}
    \item Attribute inference attack: In attribute inference attacks, an attacker tries to reconstruct the original training data or extract sensitive information from the model's parameters. This type of attack is particularly concerning when the model is used to process private or personal data.

    \item Re-identification Attacks: Even if individual data points are anonymized, attackers could use auxiliary information from other sources to link seemingly anonymous data back to specific individuals, thereby breaching their privacy.
    
    \item Membership Inference Attacks: These attacks attempt to determine whether a specific data point was used during the training process. By analyzing the model's outputs, an attacker can infer whether a particular data point was part of the training dataset, potentially revealing sensitive information about the data owner.

    \item Side-Channel Attacks: These attacks focus on exploiting unintended information leakage during the execution of privacy-preserving algorithms. By analyzing execution time, memory usage, or power consumption, attackers might gain insights into the data being processed.
    
\end{itemize}

\subsubsection{Inadvertent Threats}
\begin{itemize}
    \item Data Leakage: Confidentiality breach in agricultural data refers to a situation where sensitive or private information related to agriculture is unintentionally disclosed to unauthorized individuals or entities. This breach could involve the exposure of data such as farming practices, crop yields, land ownership details, financial information, and other sensitive information that farmers or agricultural organizations would prefer to keep confidential.
    \item Configuration Error: A configuration error as an unintentional error in agricultural data refers to a mistake or oversight made during the setup or arrangement of technology systems and software used in the agricultural context. These errors occur due to misconfigurations or incorrect settings that lead to unintended consequences or outcomes in the collection, storage, processing, or sharing of agricultural data.
    
    For instance, in precision agriculture where data-driven technologies are used to optimize farming practices, a configuration error could involve setting incorrect parameters for sensors, drones, or automated machinery. This might result in inaccurate data collection, leading to flawed decisions about irrigation, fertilization, or pest control.
    \item Inaccurate Encryption: Inaccurate encryption as an unintentional error in agricultural data refers to the incorrect implementation or utilization of encryption techniques meant to secure sensitive information related to agriculture. Encryption is a process of converting data into a secure, unreadable format to prevent unauthorized access or data breaches. However, when encryption is applied improperly or inaccurately, it can lead to unintended consequences and compromise the confidentiality of agricultural data.
    
    For instance, if encryption keys (the codes required to decrypt the data) are managed inadequately, there's a risk of unauthorized parties gaining access to the decrypted information. Additionally, using weak encryption algorithms or outdated encryption methods might render the data susceptible to decryption by attackers with sufficient computing power or knowledge.
    
    Inaccurate encryption practices in agriculture data could also involve failing to encrypt all necessary data fields or overlooking specific data sources, leaving certain parts of the information vulnerable to exposure. This can be particularly concerning when dealing with sensitive data such as crop yield projections, land ownership details, or financial records. 
    
    \item Unauthorized Access: Unauthorized access as an unintentional error in agricultural data refers to situations where individuals or entities gain entry to sensitive agricultural information without proper authorization or permission. This error occurs due to vulnerabilities in data security measures, misconfigured access controls, or inadvertent lapses in safeguarding agricultural data.
    
    For instance, if an agricultural database containing information about crop yields, pricing strategies, or proprietary farming techniques lacks proper access controls, unauthorized individuals could gain access to this information. This could happen due to weak passwords, lack of encryption, or overlooking permissions that restrict data access to only authorized personnel.
    
    Unintentional unauthorized access might also occur if a legitimate user inadvertently shares login credentials or access links with unintended parties or if attackers exploit a software vulnerability to gain entry to sensitive agricultural databases.
    
\end{itemize}

\subsection{Privacy Regulations in Agriculture}
\label{rule}

Code of Conduct describes contractual relations and guides the use of agricultural data, particularly on the rights of use and access of the data. Code was a collaborative step between massive institutions representing various enterprises producing animal fertilizers, seeds, feed, or farm machinery and agents of animal breeding organizations as well as farmer's cooperatives in the EU associated with the Council on Ethical \& Judicial Affairs (CEJA) \cite{CEJA} and Copa-Cogeca (that concentrates on young farmers up to 40 years of age). In establishing the Code, it was highlighted by the parties that ``the Code promotes the advantages of data sharing and allows agribusiness models, including agri-cooperatives and other agri-businesses, to shift into an age of digitally enhanced farming swiftly." \cite{EUCode}

\begin{figure*}[tp]
    \begin{center}
        \includegraphics[scale=0.45]{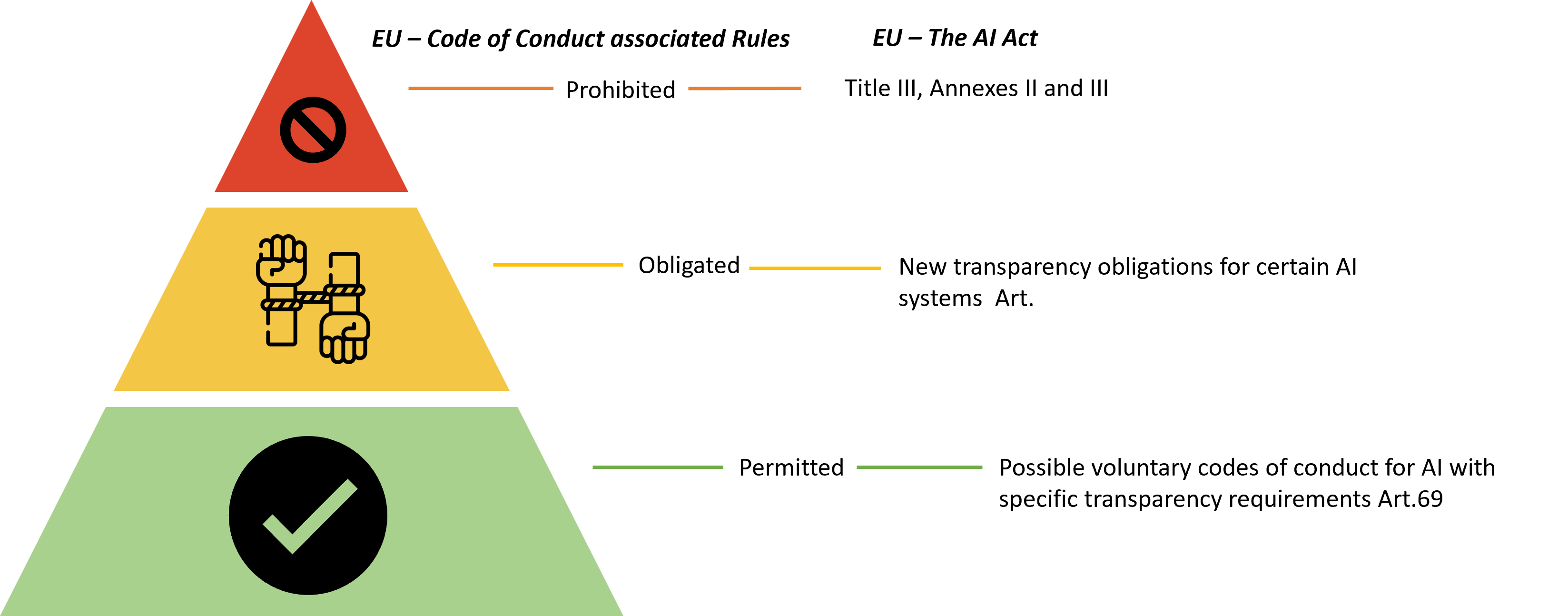}
        \caption{\centering{EU Code of Conduct Vs. AI Act}}
        \label{fig:EU_CC_Act}
    \end{center}
\end{figure*}

The European Parliament has also launched a substantial amount of work in AI. Recently, ``The AI ACT" \cite{AIACT} has been proposed to protect user's data from AI risks. The European Parliament has also launched a substantial amount of work in AI. Due to extreme technological transition in recent years and potential challenges, the EU desires balanced practices. The Union aims to preserve the EU's technological supervision and assurance that Europeans can benefit from new technologies designed and performed according to Union rules and regulations. Figure \ref{fig:EU_CC_Act} shows the relation between the EU Code and the AI Act.

\subsection{Synthetic Data in Privacy}
Various approaches have been proposed for privacy-preserving data sharing. Some common approaches towards privacy use differential privacy \cite{dwork2006calibrating}, K-anonymity \cite{el2008protecting}, L-diversity \cite{machanavajjhala2007diversity} and t-closeness \cite{li2007t}. Privacy-Preserving Generative Models use Generative models like Generative Adversarial Networks (GANs) to generate synthetic data that closely mimics the original data distribution. This synthetic data can be shared without revealing actual individual data thus providing privacy.

Generative adversarial Networks (GANs) is a class of deep learning-based generative model in AI. GANs are extremely accurate in synthetic data generation and translation, particularly for image and text data \cite{brock2018large, isola2017image, zhang2017stackgan, wang2018high}. The principal architecture in a GAN framework involves a generative model G that captures the data distribution, and a discriminative model D that estimates the probability that a sample came from the original distribution rather than G. The training procedure for G is to maximize the probability of D making a mistake. As the training progresses, the generator gets better at generating new examples that plausibly come close to the samples from the original distribution. The idea behind GAN can be formulated as a two-player min-max game with value function $V (G, D)$:
\begin{equation}
    \begin{aligned}
        {\underset{G}{\min}}\,{\underset{D}{\max}}\, V(G,D)= E_{x \sim p_{data}(x)}[log D(x)] 
        \\ + E_{z \sim p_z(z)}[log(1 - D(G(z)))]
    \end{aligned}
\end{equation}

For agricultural data, the data is usually tabular i.e. a mix of discrete and continuous values. Additionally, the continuous values are not arbitrarily random and usually follow a specific distribution within a given range. To account for this, we need specialized versions of GAN that can accurately replicate system data that is collected over our digital twin. In an unconditioned generative model, there is no control over the modes of the data being generated. Conditional Generative Adversarial Nets (CGAN) \cite{mirza2014conditional} introduces the concept that by conditioning the model on additional information, it is possible to direct the data generation process. The objective function of the two-player minimax game is rewritten as:
\begin{equation}
\begin{aligned}
    {\underset{G}{\min}}\,{\underset{D}{\max}}\, V(G,D) = E_{x \sim p_{data}(x)}[log D(x|y)]
    \\ + E_{z \sim p_z(z)}[log(1 - D(G(z|y)))]
\end{aligned}
\end{equation} 

To ensure that the synthetic dataset is distributionally close to the original dataset, and provides privacy protection by the principle of t-closeness, Kotal et al. \cite{kotal2022privetab} propose the use of Earth Mover's distance (EMD). he The EMD of the distribution of features in the synthetic is calculated w.r.t. the original dataset. The sampling process continues to sample from the trained generator till the generated distribution is within a threshold distance of the original distribution. To address the challenges of tabular data, the model uses three key steps during generation: (1) Mode-specific normalization, (2) Conditional Generator, and (3) Training by sampling.

\begin{figure*}[tp]
    \begin{center}
        \includegraphics[width=0.9\textwidth]{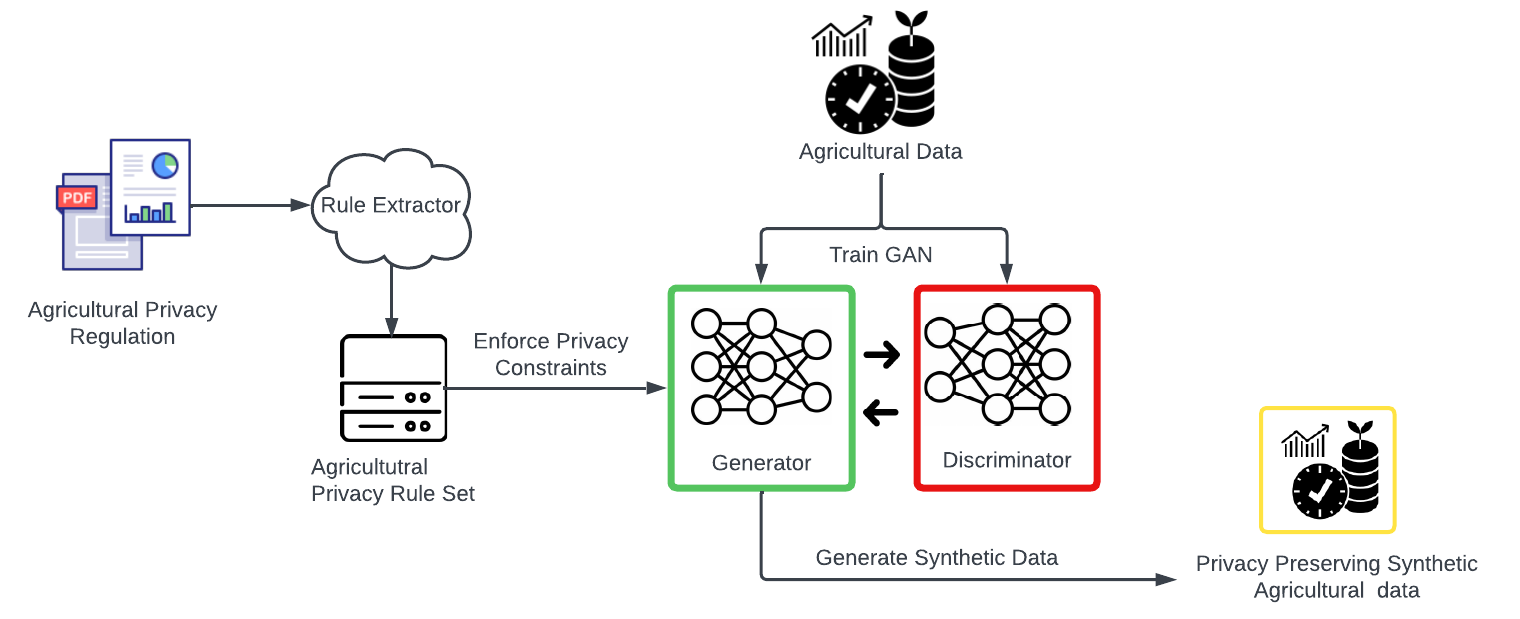}
        \caption{\centering{Framework for Privacy-Preserving Data Generation in Agriculture with Privacy Policy Enforcement}}
        \label{fig:overall_arch}
    \end{center}
\end{figure*}

\section{Privacy Preserving Data Generation with Policy Enforcement}
\label{sec_framework}
The various models proposed for privacy-preserving data sharing do not take into regulations on Information Privacy. While they can provide privacy protection for data, there is still a gap between privacy as ensured by these models and privacy as required by the law. Bridging this gap requires translating legal privacy requirements into machine-enforceable values and mechanisms. This involves understanding the key principles of data protection regulations and incorporating them into designing and implementing privacy-preserving models. This work proposes a novel framework for Policy Enforcement in privacy-preserving data generation. We process the relevant privacy regulations and extract privacy rules from the human written document. The rules are converted into a machine-readable and enforceable format. The privacy rule sets are then used to guide the privacy-preserving data generation model. Figure \ref{fig:overall_arch} demonstrates the overall architecture of our proposed framework.

\subsection{Rule Extraction from Privacy Regulations}
\label{subsec_regulation}
We use a Deontic Logic Rule-based approach to extract relevant rules from Privacy regulations into a machine-enforceable format. First, we create a predefined list of modal verbs used to express obligation types in legal domains. Next, we tokenize each token in the set and use the Python library to get dependency analysis and POS (part of speech) tags. The algorithm searches for predefined triggers within a given sentence to extract its position (starting index), each mention of the trigger, and its dependency tags. Deontic modality \cite{ML} \cite{elluri2021bert} is predominantly used in legal documents to describe vendors' prohibitions, obligations, and permissions. For instance, `may', `must', and `must not', frame  `permission', `obligation', and `prohibition' respectively. Table \ref{tab:deontictaxonomy} defines all the deontic types, and Table \ref{tab:modalverbs} has all the model verbs associated with each deontic type. Modality guides the linguistic ability to express alternative forms the world could be and is generally represented by modal auxiliaries such as must, can, shall, will, and may. Below are some examples from the EU Code of Conduct for agriculture.

Permission Rule 1: The farmer can   provide data to land owners, potato processors, the  government, paying authorities, etc. \cite{EUCode}

Permission Rule 2: Unless otherwise agreed in the contract, the data originator can transmit this data to another data user. \cite{EUCode}

Obligation Rule 1: Contracts must not be amended without the prior consent of the data originator.\cite{EUCode}

Obligation Rule 2: Parties may not use, process, or share data without the consent of the data originator.\cite{EUCode}

Prohibition Rule 1: Data cannot be owned in the same way as physical assets. \cite{EUCode}

Prohibition Rule 2: Parties may not use, process, or share data without the consent of the data originator. \cite{EUCode}

\begin{table}[!ht] 
\renewcommand{\arraystretch}{1}
\centering  
  \caption{Deontic Types Taxonomy}\label{tab:deontictaxonomy} %
\begin{tabular}{p{2cm} p{5cm}} \toprule
    {Deontic Modality} & {Definition}  \\ \midrule
    Permission  &  Vendor is authorized to do something. \\
    Prohibition  & Vendor is forbidden to do something.      \\
    Obligation  & Vendor is mandated to do something.\\
    Entitlement  & Vendor has the right to do something. \\ \midrule
\end{tabular}
\end{table}

\begin{table}[!ht]
\renewcommand{\arraystretch}{1}
    \centering
      \caption{Modal verbs for Deontic Types}\label{tab:modalverbs} %
    \begin{tabular}{| p{1.5cm} | p{4cm}|}
    \hline
        Type & Model Verbs \\ \hline
        Obligation & shall be required, will be required, shall be obligated, shall, must, will, have to, should, ought to have, will be paid, shall be paid, agree, agrees, acknowledges, acknowledge, represents and warrants, shall be responsible for, will be responsible for \\ \hline
        Prohibition & shall not, will not, must not, may not, cannot, shall have no right, can not, shall not be allowed, will not be allowed, shall not assist, shall be prohibited, will be prohibited, nor shall, not to be, neither lessor nore lessee may, in no event shall, nor will, will not allow, nor may \\ \hline
        Permissions & shall be permitted, shall also be permitted, can, may, could, shall be allowed, will be allowed, is permitted, will allow, has the right, or at landlord's option, shall be permitted to \\ \hline
    \end{tabular}
\end{table}

\subsection{Policy enforcement in Data Generation}
The rules extracted from policy documents in Section \ref{subsec_regulation} form a rule set. This rule set is used to guide the data generation process. We use the Privetab model \cite{kotal2022privetab} for privacy-preserving data generation. The principle of privacy used in the model is t-closeness \cite{li2007t} which stipulates that the distributional similarity between the original and synthetic data should be within a threshold to ensure privacy. The threshold value is not pre-determined and often determined based on the needs of an organization. To ensure that these privacy constraints meet the requirements of the privacy regulations, we use two key steps: 
\subsubsection{Determine Attribute Sensitivity} According to the EU Code of Conduct and ``The AI Act", all the features in the dataset are associated with the relevant rule and risk level. We use the rule set extracted from the privacy policy to determine the risk level of each attribute in the original data. Attributes are categorized into three levels of sensitivity: \textbf{low}, \textbf{medium} and \textbf{high}. Highly sensitive attributes are extremely risky to share and thus require the highest privacy protection. Attributes with low sensitivity are less risky and thus can be shared with minimal risks. 
\subsubsection{Enforce privacy threshold based on Sensitivity category} When generating the privacy-preserving synthetic dataset, we have to ensure that the privacy threshold is satisfied. The value of the privacy threshold can be tuned. Based on the attribute sensitivity of the dataset determined from privacy regulations, we modulate the privacy threshold for each attribute. The high-risk attributes have the strictest threshold for privacy. The privacy threshold for low-risk attributes is less strict. This ensures that the privacy requirements of the regulations are met in this process while maintaining the utility of the generated data for downstream tasks.

\section{Experimental Results}
\label{results}
Our framework for privacy-preserving data generation with policy enforcement can be used for sharing sensitive datasets. We demonstrate the use of our framework for Agricultural Big Data as a proof-of-concept. In this section, we use our framework to generate a privacy-preserving version of the ITM4Impact dataset \cite{ITM4Impa86:online}. The implementation of our policy-enforced privacy-preserving data generation framework is made available in a library\footnote{\url{https://github.com/Ebiquity/policy_enforced_data_generation}}. We provide the experimental results to demonstrate that our framework provides a privacy-preserving alternative for sensitive datasets that is still useful in downstream tasks. 

\subsection{Dataset}
In this experiment, we use the ITM4Impact dataset \cite{ITM4Impa86:online} collected by the ILRI institute to measure and determine the impact of infection and treatment method (ITM) on farmers. This dataset was anonymized for public release. We observe that some of the information shared in this dataset is privacy sensitive and poses a risk towards re-identification of participating farmers. In the ITM4Impact dataset, there are around 800 attributes. Approximately more than 140 groups segregate these attributes into multiple categories. According to the EU Code of Conduct and ``The AI Act", all the features in the dataset are associated with the relevant rule and risk level. More than 500 attributes are related to the farmer PII data or the data generated in the farming environment and categorized with a ``High" privacy level. More than 270 attributes, as per regulation, fall under ``Medium" risk, and only around 30 features can be public and can be shared with the public. However, we see that many breaches have happened often, and people end up paying huge penalties for not securing the information by adhering to the regulation policies. As mentioned in the regulatory document, it is crucial that this data should be secured and cannot be shared unless the individual has permission. 

Below are some of the rules extracted from the regulation for applying them to the relevant attributes in the dataset:\\
The data originator can store data in a primary location, in a data platform, or cloud-based storage platforms.\cite{EUCode}\\
The datasets should only be kept for as long as is strictly necessary for the relevant analyses to be carried out.\cite{EUCode}\\
If the data is being used to make decisions about the data originator “as a natural person” the GDPR applies. For instance, the rights regarding data
produced on the farm or during farming operations are granted to (“owned by”) the farmer and may be used extensively by them.\cite{EUCode}

\subsection{Fidelity Results}
Evaluating the fidelity of synthetically generated data is a critical step in assessing the effectiveness of your privacy-preserving data generation framework. By measuring the realisticness of synthetic data, we can determine how closely it resembles real data and its potential for utility in downstream tasks. We provide two fidelity metrics for our model.

\subsubsection{T-SNE Visualization} For utility in downstream tasks, it is important that the synthetic dataset provides the same data coverage as the original dataset. As observed in Figure \ref{fig:tsne} for privacy-preserving data generation, there is minimal loss in coverage from the original dataset from both with and without policy enforcement in the generation process. 

\begin{figure}[htp]
\begin{subfigure}{\columnwidth}
\centering
  \includegraphics[width=0.85\columnwidth]{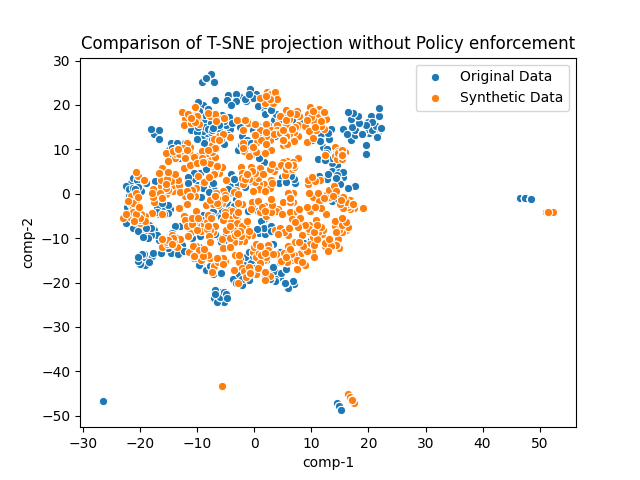}
  \label{fig:tsne1}
\subcaption[]{Comparison of T-SNE projection of Synthetic Data without Policy Enforcement}
\end{subfigure}
\begin{subfigure}{\columnwidth}
  \centering
  \includegraphics[width=0.85\columnwidth]{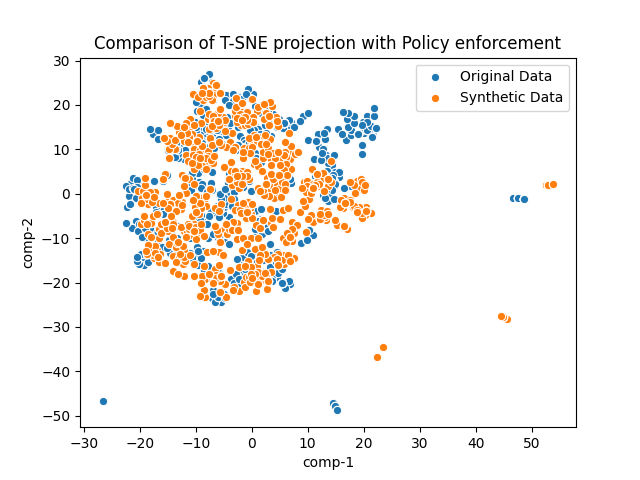}
  \label{fig:tsne2}
  \subcaption[]{Comparison of T-SNE projection of Synthetic Data with Policy Enforcement}
\end{subfigure}
\caption{Comparison of T-SNE projection of Original and Privacy-preserving Synthetic Data}
\label{fig:tsne}
\end{figure}%

\subsubsection{Statistical Similarity} We provide an of the statistical resemblance between the original and synthetic datasets for attributes at the 3 different risk levels. In general, the metrics demonstrate a strong consistency between the synthetic datasets generated with our framework and the original dataset. Figure \ref{fig:cdf_low}, \ref{fig:cdf_med} and \ref{fig:cdf_high} shows the cumulative distribution function (CDF) graphs comparing the original and synthetic data for three attributes at low, medium, and high privacy risk level respectively. The average KS statistic, which measures the maximum difference in CDFs between the original and synthetic data is 0.04 for all 3 attributes.

\begin{figure}[]
\centering
  \includegraphics[width=\columnwidth]{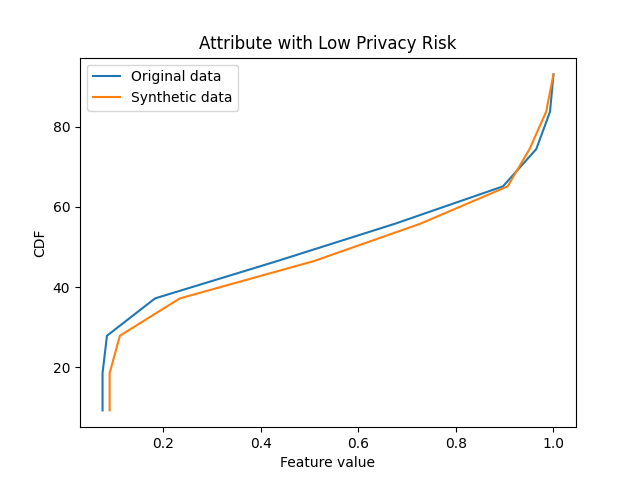}
    \caption{Comparison of CDF for Original vs Synthetic data in Low Risk Attribute}
    \label{fig:cdf_low}
\end{figure}
\begin{figure}[]
\centering
  \includegraphics[width=\columnwidth]{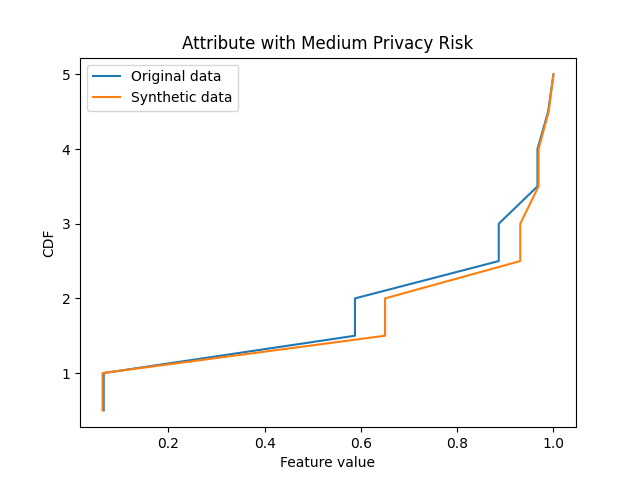}
  \caption{Comparison of CDF for Original vs Synthetic data in Medium Risk Attribute}
  \label{fig:cdf_med}
\end{figure}
\begin{figure}[]
\centering
  \includegraphics[width=\columnwidth]{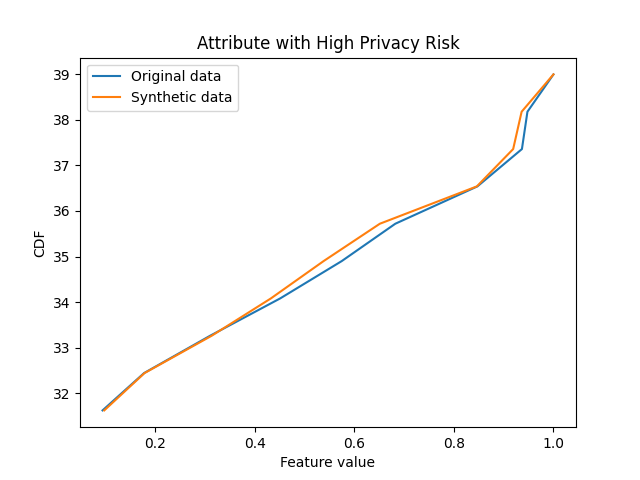}
  \caption{Comparison of CDF for Original vs Synthetic data in High Risk Attribute}
  \label{fig:cdf_high}
\end{figure}

\subsection{Utility Results}
It is important that the resultant dataset of our framework is still useful in replacing the original dataset for downstream tasks. One of the most important use cases of big data is in the field of Machine learning (ML). We show here that the data generated from our framework can replace the original dataset for ML prediction tasks. We evaluated ML models trained on the original dataset against ML models trained on the generated data. Consistent performance across both models would signify that the synthetic data retains relevant information for the task. It is important to note here the test set should be the same to compare the performance of both models. One of the use cases of ML prediction for agricultural data is farmer category prediction. We use 4 ML predictive models here: Logistic Regression (LR), Decision Tree (DT), Random Forest (RF), and Gradient Boosting Classifier (GBC). As observed in Figure \ref{fig:utility1} and Figure \ref{fig:utility2}, the ML prediction models have comparable performance on data generated both with and without policy enforcement. The average accuracy of ML models trained on the original data is 0.95. The average accuracy of ML models trained on synthetic data generated without policy enforcement is 0.9. The average accuracy of ML models trained on synthetic data generated without policy enforcement is 0.895. The average loss in accuracy for ML models trained on synthetic data generated without policy enforcement is 0.05. The average loss in accuracy for ML models trained on synthetic data generated without policy enforcement is 0.055. Thus there is an average loss of 0.005 in accuracy of data generated with vs without policy enforcement. Thus the data generated from our framework retains the most relevant information for downstream ML tasks. 

\begin{figure}[htp]
\begin{subfigure}{\columnwidth}
\centering
  \includegraphics[width=\columnwidth]{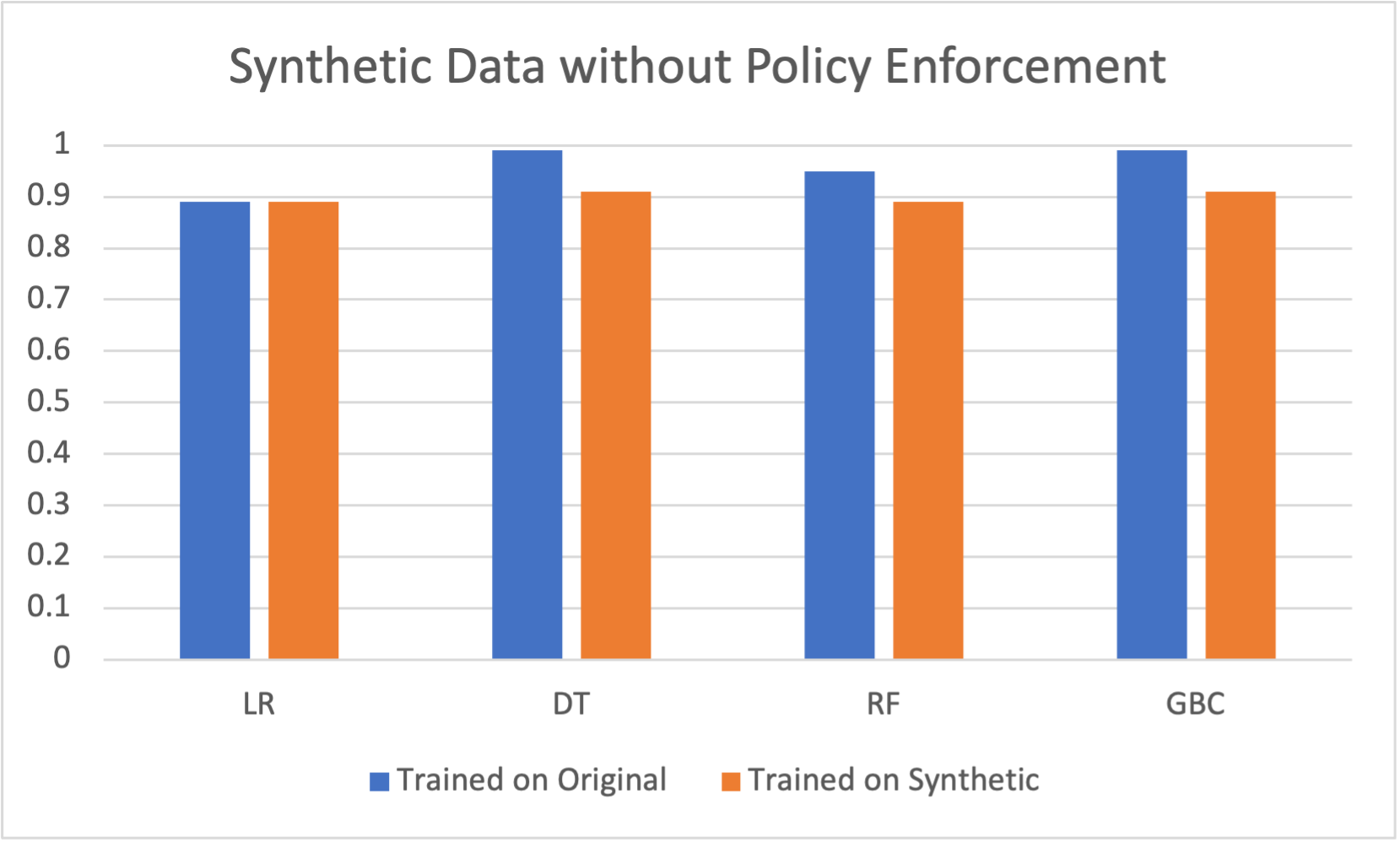}
  \subcaption[]{Synthetic Data without Policy Enforcement}
  \label{fig:utility1}
\end{subfigure}
\begin{subfigure}{\columnwidth}
  \centering
  \includegraphics[width=\columnwidth]{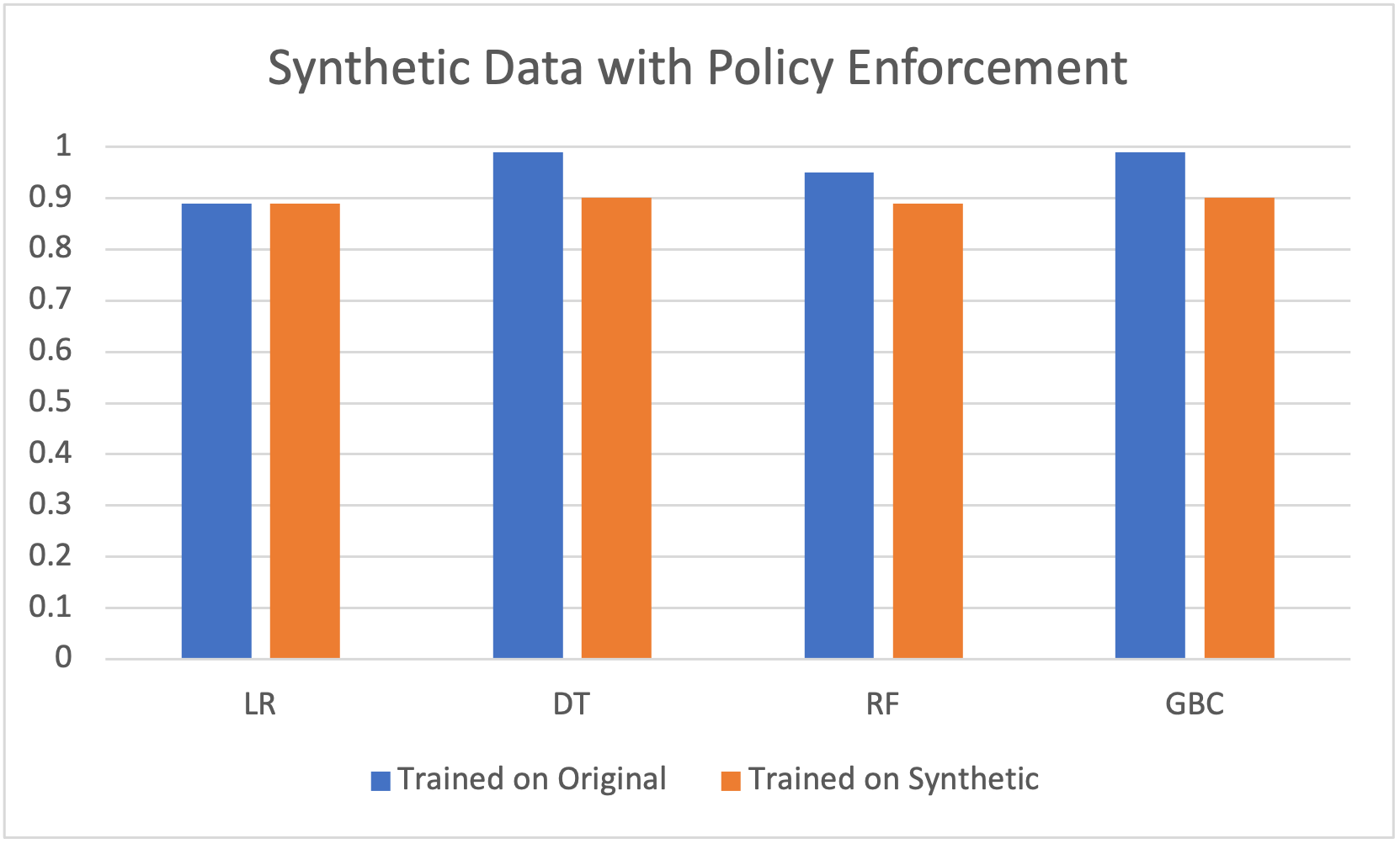}
    \subcaption[]{Synthetic Data with Policy Enforcement}
    \label{fig:utility2}
\end{subfigure}
\caption{Prediction performance with the model trained on Original vs Privacy-preserving Synthetic Data}
\label{fig:utility}
\end{figure}%

\subsection{Privacy Results}

\subsubsection{Attribute Inference Attack} In attribute inference attacks, an attacker tries to reconstruct the original training data or extract sensitive information from the model's parameters. In this setting, an adversary has partial knowledge of some training records and access to a model trained on those records and infers the unknown values of a sensitive feature of those records. Attribute inference attacks typically use ML models to learn about the original dataset from the predictions of the trained model and try to predict the values of unknown sensitive attributes with a confidence score. Efficient Attribute Inference attacks can break privacy-preserving efforts by unmasking the values of sensitive attributes. The lower accuracy of attribute inference attack models proves that privacy-preserving algorithms are efficient in evading these attacks. In Figure \ref{fig:attrib_infer}, we compare the results of a re-identification attack against privacy-preserving datasets generated with and without policy enforcement. For all 3 risk levels, the accuracy of the attribute inference attack model is lower for the dataset generated with our framework of policy enforcement. for privacy-preserving data generated without policy enforcement, the average attribute inference attack accuracy is 0.35 across the 3 privacy risk levels. For privacy-preserving data generated without policy enforcement, the average attribute inference attack accuracy is 0.3 across the 3 privacy risk levels. Thus, there is a 0.05 increase in accuracy loss for data generated with policy enforcement. For attributes with high privacy risk, the increase in accuracy loss is 0.08. This shows that the dataset generated with our approach is more resistant to attribute inference attacks.

\begin{figure}{}
  \centering
  \includegraphics[width=\columnwidth]{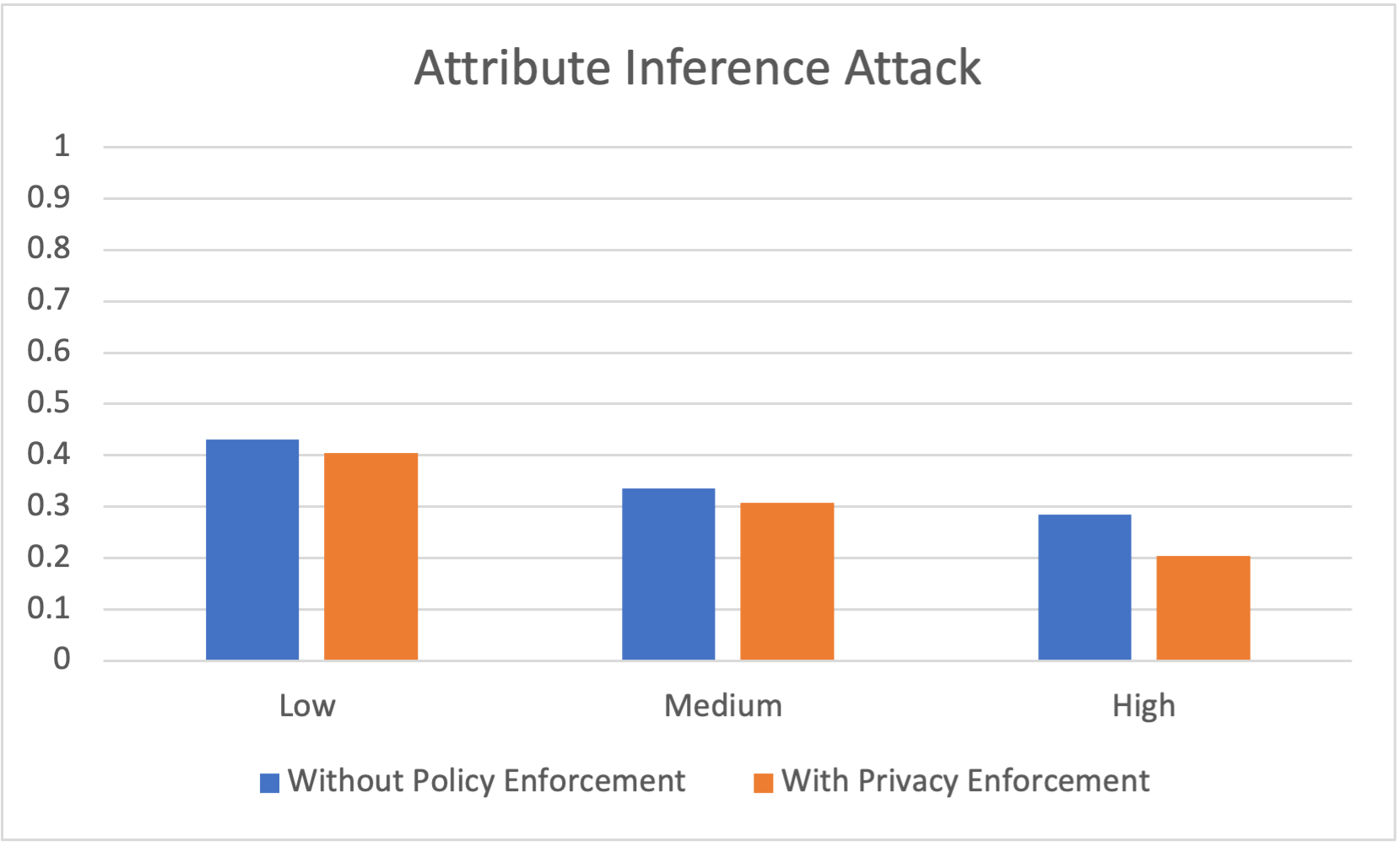}
  \caption{Comparison of Accuracy of Attribute Inference attacks against privacy preserving data generation with vs without policy enforcement}
  \label{fig:attrib_infer}
\end{figure}

\subsubsection{Re-identification Attack} Re-identification attacks are variations of linkage attacks where even if individual data points are anonymized, attackers could use auxiliary information from other sources to link seemingly anonymous data back to specific individuals, thereby breaching their privacy. Efficient re-identification algorithms use ML to infer information about anonymized datasets with partial knowledge of the original dataset or auxiliary information. It is then used to predict de-anonymized values of sensitive attributes in a dataset. Higher accuracy of re-identification attacks in de-identified datasets shows that the dataset still contains considerable privacy risk. In Figure \ref{fig:reident}, we compare the results of the re-identification attack against privacy-preserving datasets generated with and without policy enforcement. Data generated without policy enforcement has a consistent accuracy against re-identification attacks for the 3 risk levels of attributes ($\approx 0.50$). For attributes with low privacy risk, the re-identification attack model has comparable accuracy ($0.51$) against our framework with policy enforcement. However, for attributes with medium and high privacy risk, our framework with policy enforcement works better with an average accuracy loss of 0.07 in re-identification attacks. Thus the dataset generated with our approach is more resistant against re-identification attacks.

\begin{figure}{}
  \centering
  \includegraphics[width=\columnwidth]{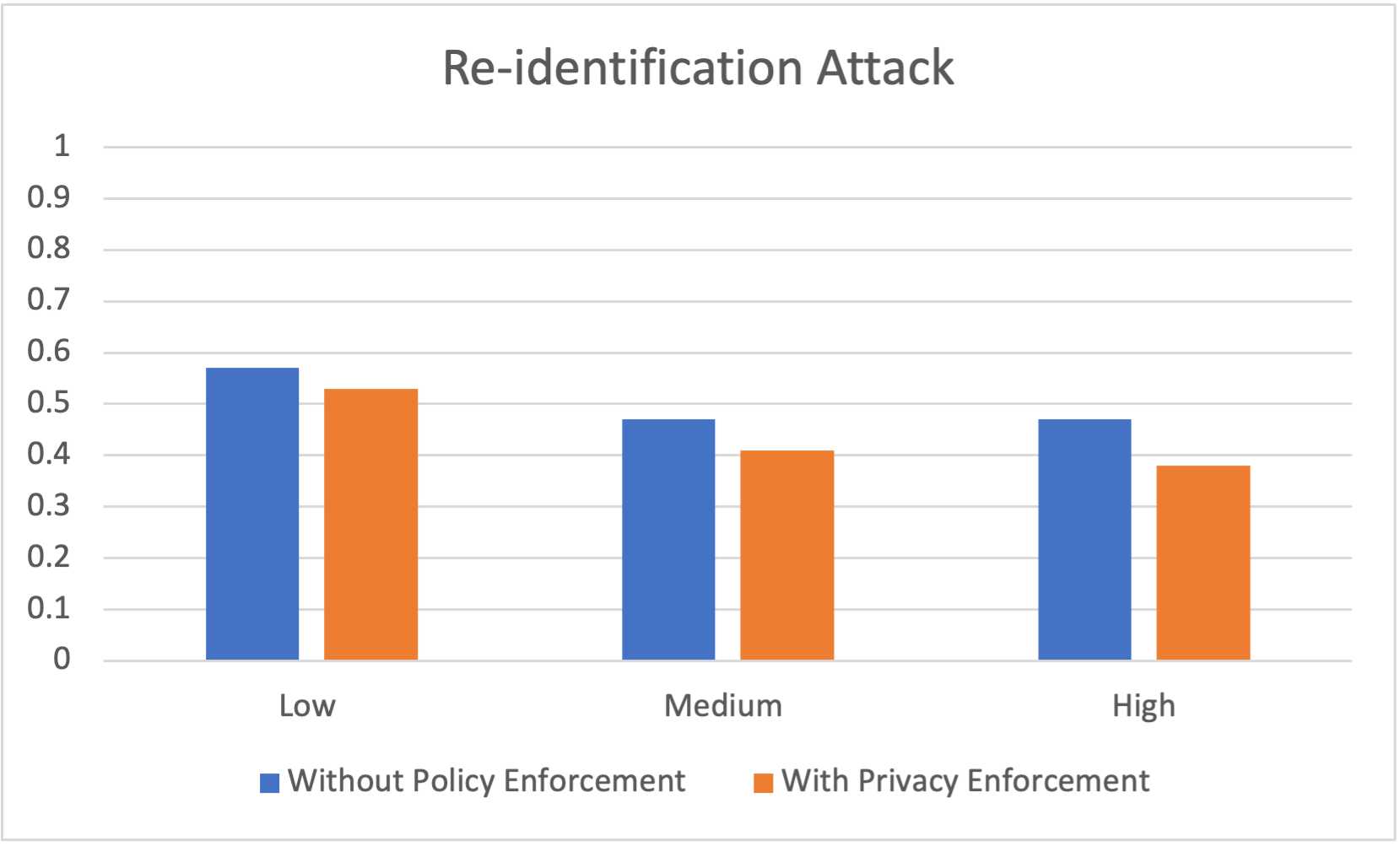}
  \caption{Comparison of Accuracy of Re-identification attacks against privacy preserving data generation with vs without policy enforcement}
    \label{fig:reident}
\end{figure}

\section{Related Work}
\label{related}
In the absence of precisely elaborated rules on the appropriate ways to deal with the legal implications of digital farming technologies on agricultural companies and relationships, farmers and their agribusiness organizations who deliver digital farming technologies started to shape their policies to enhance agricultural data management techniques and provide a foundation for reliable data sharing. Agricultural security and privacy principles and codes of conduct have been developed in different parts of the world. In 2014, in the U.S., the American Farm Bureau's Privacy and Security Principles for Farm Data (2019) \cite{AFBF} was the foremost to draw awareness to some of the problems farmers had about how their information was being collected and shared. After this initiative was 2014 New Zealand's Farm Data Code of Practice \cite{NZFD}, and more recently, in 2019, we noticed the E.U. Code of Conduct launch on agricultural data sharing by contractual agreement \cite{EUCode}. 

With the recent advances in Big Data, there is a gap between the computational capabilities and the data available for research. Data scientists are invested in gathering large volumes of data with secure and privacy-preserving approaches. Privacy-preserving methods and the impact of privacy policies on data sharing have been an ongoing avenue of research \cite{kotal2020vicloud, elluri2020measuring, kotal2021effect, elluri2021policy, echenim2023ensuring}. There has been significant research in designing privacy-preserving data-sharing methods. However, most approaches have a caveat associated with them. Among these various privacy approaches, differential privacy \cite{dwork2006calibrating}, K-anonymity \cite{el2008protecting}, L-diversity \cite{machanavajjhala2007diversity}, and  t-closeness \cite{li2007t} are worth noting. It has also been shown that achieving one of the privacy metrics can ensure others \cite{domingo2015t}.

Synthetic data can provide anonymity to original data without loss of accuracy in downstream data analysis tasks.  In general, there is a lot of evidence of GANs being used for synthetic data generation and translation in image and text data \cite{brock2018large, isola2017image, zhang2017stackgan, wang2018high}. However, the properties of system or device data make it distinct from image and text data.  The system data is usually tabular i.e. contains a mix of continuous and discrete variables, and in some cases, the sequence of consecutive rows in the data is important. A conditional generator model can address the issue of mixed attributes in tabular data by seeking to minimize the distance between generated and real data given a fixed value of the discrete variable \cite{arjovsky2017wasserstein, xu2018synthesizing, xu2019modeling, kotal2022privetab, das2023change, piplai2023knowledge}.   

While there are privacy-preserving models for data sharing, these models do not take into account the rules stated in privacy policies such as GDPR \cite{elluri2018integrated} \cite{elluri2018knowledge}, EU code of conduct \cite{EUCode} etc. Hence, there is no mechanism to ensure that data shared through these frameworks are compliant with the privacy policies. To bridge this gap, we propose a novel framework that learns rules from the privacy policies and inducts the information in the generation process of a privacy-preserving GAN to ensure that the data generated is privacy-preserving, secure to share, and compliant with the data policies.

Gupta et al.~\cite{gupta2020security} presented a vast exposure to cybersecurity threats and vulnerabilities in smart farming environments. This research ~\cite{makhdoom2020privysharing} proposed PrivySharing, a blockchain-based innovative framework for privacy-preserving and secure IoT data sharing in a smart city environment. West~\cite{west2018prediction} introduced a principles-based framework to assess cyber-attack vulnerabilities and also constructed a precision agriculture system protected from cyber-attacks. Coble et al.~\cite{coble2018big} discussed a set of analytical techniques that are increasingly relevant to solving security and privacy issues. Kumar et al.~\cite{kumar2021sp2f} proposed a Secured Privacy-Preserving Framework (SP2F) for smart agricultural Unmanned Aerial Vehicles (UAVs), which handles various cyber attacks. 

In addition, several security models for protecting big data in various domains are discussed in~\cite{gupta2020access, gupta2021future, aslan2021intelligent, ozkan2021comprehensive, gupta2021detecting, gupta2021hierarchical, gupta2022game}.

\section{Conclusion and Future Work}
\label{conclusion}
Technology, particularly the integration of big data analysis, data processing, cloud computing, and IoT devices, plays a pivotal role in enhancing agricultural output both quantitatively and qualitatively. Utilizing these technological advancements enables the agricultural sector to address intricate challenges by effectively analyzing large volumes of crowd-sourced data. However, the sharing of data across various sources raises privacy concerns. While GDPR and similar general privacy regulations provide a foundation for digital data sharing, the specific privacy risks pertinent to agricultural data remain unclear. Defining agricultural big data is still an ongoing debate, making it challenging to delineate privacy concerns. This study delves into the EU Code of Conduct on agricultural data sharing, established through contractual agreements as an extension of GDPR and the E.U. A.I. Act, to comprehensively understand privacy concerns in the realm of agricultural data. Though there have been privacy-preserving efforts in data sharing, there is a lack of policy enforcement in such efforts. 

To address the challenge of data sharing while ensuring consistency with privacy regulations, this paper introduces a novel framework for enforcing policy rules in privacy-preserving data sharing. This approach not only mitigates privacy risks like leakage and re-identification but also bridges the gap between privacy-preserving methods and policy-defined privacy requirements. The experimental results show that this framework retains substantial statistical similarity with original data and retains utility in downstream tasks while being resistant to privacy threat models. In conclusion, the proposed framework showcases a promising direction for reconciling agricultural data sharing and privacy concerns, ensuring a sustainable and secure future for agricultural research and advancement. In the future, we would like to extend this framework for policy enforcement in other domains that need privacy-preserving efforts in data sharing, such as health, security, etc. 

\bibliographystyle{IEEEtran}
\bibliography{refs}

\end{document}